\documentclass[pre, preprint, groupedaddress,showpacs]{revtex4}
\usepackage{graphicx}
\begin{document}

%\preprint{}

\title{Enhancement of the longitudinal transport by a weakly transversal drive}


\author{Ya-feng He$^{1,2}$}\email[Email: ]{heyf@hbu.edu.cn}
 \author{Bao-quan  Ai$^{1}$}
%\homepage[]{}

%\thanks{}
\affiliation{$^{1}$Laboratory of Quantum Information Technology,
ICMP and SPTE, South China Normal University, 510006 Guangzhou, China.\\
 $^{2}$College of Physics Science and Technology, Hebei University, 071002 Baoding, China}


\date{\today}
\begin{abstract}
\indent Transport of Brownian particles in a two-dimensional
asymmetric tube is investigated by applying a polarized field. From
the Brownian dynamics simulations we find that the longitudinal
current can be enhanced remarkably by applying a weakly transversal
drive.  Multiple current reversals can be realized by altering the
driving frequency of the polarized field. By coupling the
longitudinal and transversal forces together, one can control the
particle transport flexibly.
\end{abstract}

\pacs{ 05. 40. -a, 05. 60. -k, 07. 20. Pe}
%\keywords{periodic tube, polarized field, multiple current
%reversals}



\maketitle
\section {Introduction}
\indent The problem of non-equilibrium-induced transport processes
has attracted much interest in theoretical as well as experimental
physics \cite{Julicher,Hanggi,Reimann1,Rousselet}. This subject was
motivated by the challenge to explain unidirectional transport in
biological systems \cite{Julicher}, as well as their potential
technological applications ranging from classical non-equilibrium
models \cite{Rousselet,Faucheux} to quantum systems
\cite{Derenyi,Lee}. In these systems directed-Brownian-motion of
particles is generated by non-equilibrium noise in the absence of
any net macroscopic forces and potential gradients. Ratchets
\cite{Magnasco,Reimann2,Reimann3,Doering} have been proposed to
model the unidirectional motion due to the zero-mean non-equilibrium
fluctuation.

\indent Recently, ratchet effects were realized in many experimental works \cite{Savel'ev, Cole, Villegas, Clecio, Kalman, Gommers}. Savel'ev and Nori \cite{Savel'ev} proposed devices for controlling the motion of flux quanta in superconductors and could address a central problem in many superconducting devices. They also demonstrated experimentally how to guide flux quanta in layered superconductors by using a drive that is asymmetric in time instead of being asymmetric in space \cite{Cole}. Once or multiple reversal in the direction of the vortex flow has been observed \cite{Villegas, Clecio}. Motivated by the theoretical study in Ref. \cite{Savel'ev2}, Kalman and co-workers \cite{Kalman} studied ion current rectification in single conical nanopores in polymer films in the presence of two rectangular voltage signals. In addition, the ratchet effects were also realized in cold atoms. Renzoni and co-workers \cite{Gommers} demonstrate experimentally a gating ratchet with cold rubidium atoms in a driven near-resonant optical lattice. A single harmonic periodic modulation of the potential depth is applied, together with a single-harmonic rocking force. Directed motion is observed as a result of the breaking of the symmetries of the system.

\indent Most studies have revolved around the energy barriers \cite{Bier, Kostur}.
However, in many transport phenomena \cite{Chou}, such as those
taking place in micro- and nano-pores, zeolites, biological cells,
ion channels, nanoporous materials and microfuidic devices etched
with grooves and chambers, Brownian particles, instead of diffusing
freely in the host liquid phase, undergo a constrained motion.
 Recently, Reguera and coworkers \cite{Reguera} used the mesoscopic nonequilibrium
 thermodynamics theory to derive the general kinetic equation of the system and studied the current and
 the diffusion of Brownian particles moving in a symmetric channel with a biased external force.
  Ai and co-workers \cite{Ai1} studied the transport driven by the longitudinal ac forces and found that
  the motion of particles can indeed be rectified, with a sign that depends on the details of the wall profile.
  Interestingly, the net current can even be obtained by applying a transverse ac drive, like a transverse ac force
  or transverse wall vibration \cite{Marchesoni,Ai2}.

\indent However, if a polarized field coupled by longitudinal force
and transversal force is applied on Brownian particles, how about the competition and cooperation between them in the transport? In the present work, we study the particle transport in an asymmetrically periodic tube driven
 by thermal noise and polarized field by using Brownian dynamics simulation method. We emphasize on the enhancement
 and secondary reversal of the current by coupling the longitudinal and transverse forces together.

\section{Model and Methods}
\indent Because most of the molecular transport occurs in the overdamped regime, the inertial effects can be neglected in the study. So the overdamped dynamics can be described by the following Langevin equations in the dimensionless form,

\begin{eqnarray}
  \frac{dx}{dt} = f_{x}(t)+\sqrt{D}\xi_{x}(t),\\
  \frac{dy}{dt} = f_{y}(t)+\sqrt{D}\xi_{y}(t),
\end{eqnarray}
where, $x$, $y$ are the two-dimensional coordinates, D is the diffusion coefficient, and $\xi_{x,y}(t)$
presents the Gaussian white noise with zero mean and correlation
function: $<\xi_{i}(t)\xi_{j}(t^{'})>=2\delta_{i,j}\delta(t-t^{'})$
for $i,j=x, y$. $<...>$ denotes an ensemble average over the
distribution of noise. $\delta$ is the Dirac delta function. Two
periodic forces are applied on the particles in the longitudinal and
transversal directions, respectively with the forms:

\begin{eqnarray}
  f_{x}(t)=A_{x} \cos(\omega t+\phi_{1}),\\
  f_{y}(t)=A_{y} \cos(\omega t+\phi_{2}),
\end{eqnarray}
where, $A_{x}$, $A_{y}$ are the amplitudes of the forces in the
longitudinal and transversal directions, respectively. $\phi_{1}$
and $\phi_{2}$ are the phase of the sinusoidal forces. Here, we only
consider the case that the forces in the two directions have
identical frequency $\omega$. The superposition of $f_{x}(t)$ and
$f_{y}(t)$ gives rise to a polarized field rotating in two
dimensions as shown in figure 1. The mode of the polarized field is
characterized by the phase difference
$\Delta\phi=\phi_{2}-\phi_{1}$. Driven by the polarized field, the
particles collide with the asymmetric tube with different forms.

\begin{figure}[htbp]
  \begin{center}\includegraphics[width=8cm,height=4.5cm]{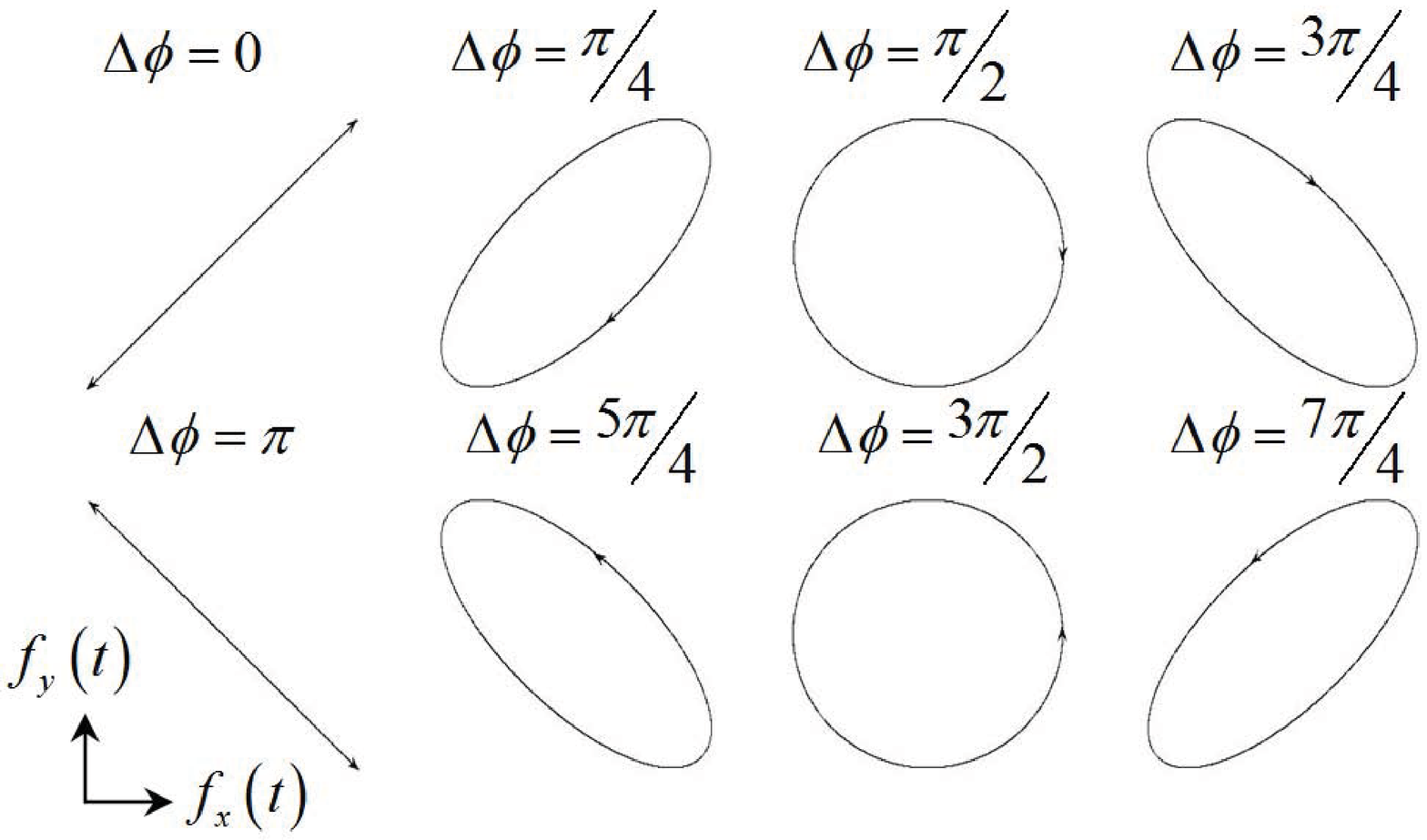}
  \caption{Polarized field by superposition of the longitudinal and transversal forces with different modes.}\label{1}
\end{center}
\end{figure}

\indent In this study we consider the case that the particles are confined in a tube.
The asymmetric tube is sketched in Fig. 2(a) with the shape described by its half width:

\begin{equation}\label{}
    y(x)=a[\sin(\frac{2\pi x}{L})+\frac{\Delta}{4}\sin(\frac{4\pi x}{L})]+b,
\end{equation}
here, $a$ is the parameter that controls the slope of the tube and
$\Delta$ is the asymmetric parameter of the tube shape. The
parameter $b$ determines the half width at the bottleneck. $L$ is
the period of the tube. The Langevin equation describing the overdamped Brownian particle can be reduced to an effective 1D Fokker-Planck equation in the absence of the longitudinal force $f_{x}$ \cite{Reguera}:

\begin{equation}\label{}
    \frac{\partial}{\partial t}P(x,t)= \frac{\partial}{\partial x}[D\frac{\partial}{\partial x}+V^{'}(x,t)]P(x,t),
\end{equation}
where, the effective potential
\begin{equation}\label{}
    V(x,t)=-D\ln[\frac{2D}{f_{y}(t)}\sinh(\frac{f_{y}(t)y(x)}{D})].
\end{equation}

\begin{figure}[htbp]
  \begin{center}\includegraphics[width=8cm,height=7cm]{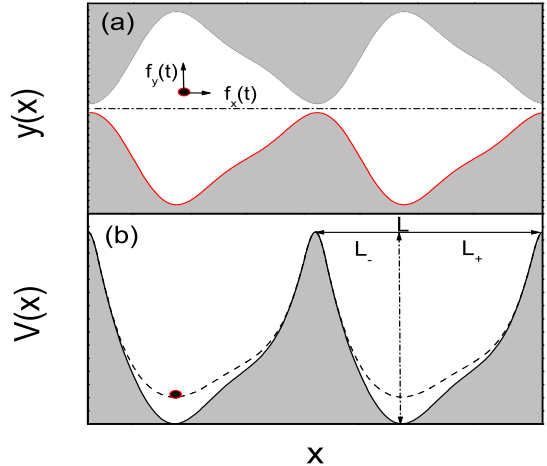}
  \caption{Schematic of the two-dimensional tube (a) and the corresponding effective potential (b). $L=L_{+}+L_{-}$ is the period of the tube, $L_{+}$ and $L_{-}$ indicate the asymmetry of the effective potential. The dash line and the solid line in (b) present the vibration range of the potential well.}\label{1}
  \end{center}
\end{figure}
\indent The transient profile of the effective potential is shown in Fig. 2 (b). The potential barriers are skewed to the right with ratchet length $L_{+}>L_{-}$. Under appropriate parameters, the maximum of the potential keeps constant, while the minimum vibrates periodically with frequency of twice of the driving frequency. This leads to periodic change on the potential barrier. Together with the longitudinal force, it gives rise to complex dynamics of the particle transport.

\indent We carry out the Brownian dynamic simulations performed by
integration of the overdamped Langevin equation using the standard
stochastic Euler algorithm. Reflecting and periodic boundary
conditions are applied in the transversal and longitudinal direction
\cite{Ai2}, respectively. The average particle velocity derived from
an ensemble average of more than $10^{4}$ along the x-direction
reads:

\begin{equation}\label{}
v=\langle\dot{x}\rangle=\lim_{t\rightarrow\infty}\frac{\langle
x(t)
    \rangle}{t}.
\end{equation}
\indent Unless otherwise noted, our simulations are under the parameters sets: $a=1/2\pi$, $b=1.2/2\pi$, $\Delta=1.0$, $D=0.1$, $\Delta\phi=0.0$, and $L=1.0$ throughout this work.

\section {Numerical results and discussion}
\indent We firstly study the determination of the frequency of the
polarized field on the particle transport. Figure 3 presents the
dependence of the mean velocity on the driving frequency for the
case that longitudinal force, transversal force, and linearly
polarized field are applied, respectively. It exhibits interesting
phenomena of enlargement and secondary reversal of current
originated from the polarized field as increasing the driving
frequency. We find that it actually demonstrates the cooperative
relationship between the longitudinal and transversal forces. In
order to illustrate this, it is necessary to discuss the cases that
the longitudinal and transversal forces are applied individually.
\begin{figure}[htbp]
  \begin{center}\includegraphics[width=8cm,height=6cm]{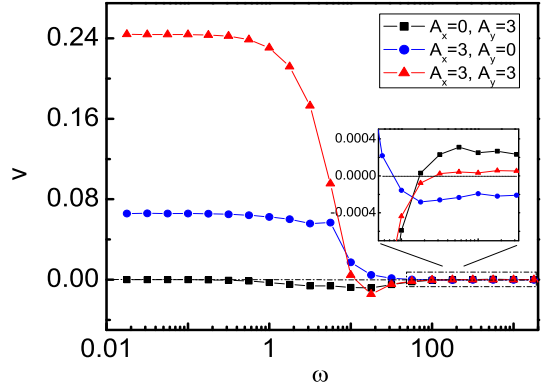}
  \caption{(Color online) Dependence of the mean velocity on the driving frequency. The square, circle, and angle lines present the applications of transversal force, longitudinal force, and polarized field. The inset shows the mean velocity at high frequency.}\label{1}
\end{center}
\end{figure}

\indent In figure 3 the circle line presents the mean velocity
induced only by the longitudinal force, $A_{x}\neq0$, $A_{y}=0$. It
can be seen that the current changes direction once as increasing
the frequency. In the adiabatic limit $\omega\to0$, the driven
source can be divided into two opposite static forces $A_{x}$ and
$-A_{x}$. The mean velocity can be expressed as
$v=\frac{1}{2}[v(-A_{x})+v(A_{x})]$. In this case the particles get
enough time to cross the both sides and it is easier for a particle
to climb the potential well to the right than to the left, which
gives rise to a positive current. As increasing the frequency, the
duration which the force acts on the particles decreases in every
period. The diffusion of the Brownian particles to the left
bottleneck depresses the particles to climb to the right wall, which
leading to a negative net current. For the extreme case
$\omega\rightarrow\infty$, the particles cannot respond the applied
force, which makes the current tend to be zero. So, the current
changes direction from positive to negative in the absence of the
transversal force.

\indent When the Brownian particles are driven periodically only in
transversal direction, $A_{x}=0$, $A_{y}\neq0$, the behaviors of the
current change dramatically as shown by the square line in Fig. 3.
In the adiabatic limit $\omega\to0$, the mean velocity also is
$v=\frac{1}{2}[v(-A_{y})+v(A_{y})]$. It is obvious that no net
current occurs when a transversal constant force is applied,
namely, $ v(-A_{y})=v(A_{y})=0$. Therefore, the current tends
to zero in the adiabatic limit.  On increasing the frequency, the
Brownian particles have enough time to diffuse in the whole area. It
has higher probability to collide with the right wall because the
area of the right wall is larger than that of the left wall, which
gives rise to a negative current. When the frequency is higher, the
particles have no enough time to reach the whole area of the right
wall. The probability of the particles colliding with the right wall
is smaller than that with the left wall, leading to a positive
current. For the extreme case $\omega\rightarrow\infty$, the
Brownian particles cannot respond the driven force, which results in
a zero current. So, the current changes direction once from negative
to positive in the absence of the longitudinal force.

\indent However, if a polarized field is applied, novel phenomena
appear as indicated by the angle line in Fig. 3. One of the
important results is that the current is enlarged remarkably when
the driving frequency is not too high. For the positive current, the
magnified effect becomes more evident. The current induced by a
polarized field is not a simple algebraic sum of those from
longitudinal and transversal forces. The amplification factor can be
control by adjusting parameters such as the driving amplitude, their
ratio $A_{x}/A_{y}$, the period of the tube and so on. It shows that
the coupling between the longitudinal and transversal forces can
improve the rectification effect of ratchets. This is very
significant in applications such as particle separation. The other
important result is the appearance of the secondary reversal of the
current as increasing the driving frequency. As stated above, the
current induced individually by longitudinal or transversal force
varies direction only once upon increasing the frequency. However,
by coupling the longitudinal and transversal forces together, the
current can undergo reversal twice. Although the magnitude of the
current in Fig. 3 is small at high frequencies (it is still within the numerical accuracy of the simulations, see Ref.\cite{Ai2}), as will be shown
later, one can improve the current by adjusting the control
parameters. So, a polarized field coupled by longitudinal and
transversal forces provides a more efficient way to control the
particle transport, especially to realize the enlargement of the
current.

\indent As a characteristic parameter of a polarized field, the
phase difference $\Delta\phi$ affects the particle transport as
well. We find that the current varies sinusoidally with a period of
$\pi$ instead of $2\pi$ as changing the phase difference (not shown). This is
because that the tube owns mirror symmetry along the longitudinal
axis. So, it is identical for the levorotatory and dextrorotatory
polarized fields when acting on the particles. When $\Delta\phi=0,
\pi$, i.e. the longitudinal and transversal forces act on the
particles with the maximum amplitude synchronously, the current
reaches maximum. However, if the longitudinal force is out of phase
with the transversal force the current will decrease. For the cases
of circular polarizations, $\Delta\phi=\frac{\pi}{2},
\frac{3\pi}{2}$, the current is minimum.

\begin{figure}[htbp]
  \begin{center}\includegraphics[width=8cm,height=6cm]{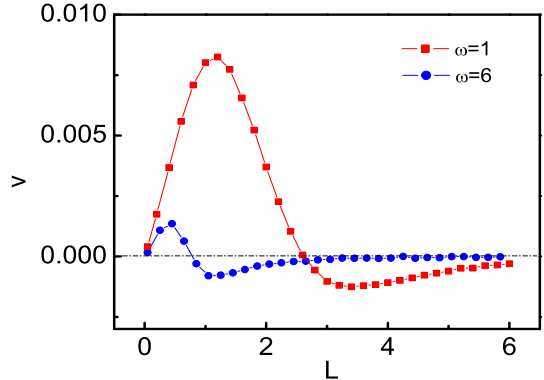}
  \caption{(Color online) Dependence of the mean velocity on the period of the tube at frequencies $\omega=1$ and $\omega=6$. The width of the tube keeps constant. The amplitudes are $A_{x}=A_{y}=0.5$.}\label{1}
\end{center}
\end{figure}

\indent There exists competitive and cooperative relationships
between the longitudinal and transversal forces on controlling the
particle transport. Figure 4 presents the dependence of the mean
velocity on the period of the tube. When the period of the tube is
very small compared with the width of the tube, the aspect ratio of
the tube $max[y(x)]/L$ is large. The particle has no enough time to
collide with the wall before it climbs the entropic barrier from the
right wall. In this case the longitudinal force predominates in
controlling the Brownian particle, which leading to a positive
current. On the contrary, if the period becomes large, the aspect
ratio is small. The particle will collide well with the right wall
before it climbs the entropic barrier from this side, which gives
rise to a negative current. Under appropriate value of period (the
cross points on the longitudinal axis), the actions of the
longitudinal and transversal forces reach balance, and the particle
couldn't escape from the well leading the net current to be zero.
For the two extreme cases, $L\rightarrow 0, \infty$, the tube tends
to be straight and loses the symmetry-breaking. So, the current
disappears. This gives us an insight that by adjusting the period of
the system, for example stretching or compressing the tube, one can
control the direction of the current.

\begin{figure}[htbp]
  \begin{center}\includegraphics[width=8cm,height=10cm]{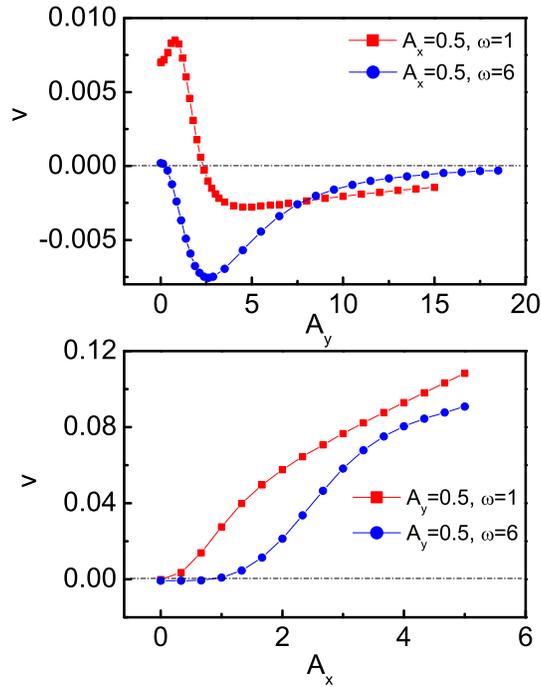}
  \caption{(Color online) Dependence of the mean velocity on the ratio of driving amplitudes at frequencies $\omega=1$ and $\omega=6$. (a) [(b)] presents the variation of the mean velocity as increasing the transversal (longitudinal) force with fixed longitudinal (transversal) force $A_{x}=0.5$ ($A_{y}=0.5$).}\label{1}
\end{center}
\end{figure}

\indent Figure 5 shows the dependence of the mean velocity on the
amplitude of $A_{y}$ and $A_{x}$, respectively. In Fig. 5 (a) [(b)],
the longitudinal (transversal) force is kept constant amplitude
$A_{x}=0.5$ ($A_{y}=0.5$) while increasing the amplitude of the
transversal (longitudinal) force. At the origin of the longitudinal
axis in Fig. 5, the current doesn't equal to zero, which is induced
by the longitudinal and transversal forces applied individually.
From figure 5 we can find that, on one hand, all the curves cross
the longitudinal axis once, which indicates the competition between
the longitudinal and transversal forces. The points of intersection
present the balance between the longitudinal and transversal actions
leading to zero current. So, by applying an addition transversal
driving force, one can further control the current direction beside
its enhancement function. On the other hand, there exist
optimization values of the amplitude at where the current reaches
absolute maxima. This illustrates the cooperative relationship
between the longitudinal and transversal forces. In Fig. 5 (b),
there should be a maximum if the longitudinal amplitude is strong
enough.
\section{Concluding Remarks}
\indent In conclusion, we have studied the transport of
overdamped Brownian particles confined in an asymmetrically periodic
tube applying a polarized field. By coupling the longitudinal and
transversal forces together, the current exhibits two types of novel
phenomena: enhancement and secondary reversal. The enhanced
efficiency and the current reversal can be controlled flexibly by
adjusting the system parameters such as the driving amplitudes,
their ratio $A_{x}/A_{y}$, the phase different, and the period of
the tube. The results we have presented may have a potential
application in many processes, such as particle separation,
diffusion in biological membrane, transport in zeolites, controlled
drug releases, and diffusion in man-made materials.

\indent Clearly, the model is too simple to provide a realistic
description of real systems. However, the results we have presented
have wide applications in may processes, such as the vortex ratchet in superconductor \cite{Savel'ev, Cole, Villegas, Clecio, Kalman}, gating ratchet with cold atoms in optical lattice \cite{Gommers}, diffusion of ions and macromolecular solutes through the channels in biological membranes \cite{c1}, transport in zeolites \cite{c2}, and diffusion in
man-made periodic porous materials \cite{Chou}.

\section{ACKNOWLEDGMENTS}
\indent  This work was supported in part by National Natural Science
Foundation of China with Grant No. 30600122, 10947166 and GuangDong Provincial
Natural Science Foundation with Grant No. 06025073. Y. F. He also
acknowledges the Research Foundation of Education Bureau of Hebei
Province (Grant No. 2009108).

\end{document}